\documentclass[fleqn,twoside]{article}
\usepackage{espcrc2}
\usepackage{graphicx}
\usepackage{epsfig}

\title{{\noindent\small
UNITU-THEP-8/04 \hspace*{1cm} IPPP/04/43  \hspace*{1cm} DCPT/04/86 \hfill 
hep-ph/0407332}\\ Semiperturbative construction for the quark-gluon vertex
}

\author{Felipe J. Llanes-Estrada \address{Universidad Complutense de 
Madrid, Depto. F\'{\i}sica Te\'orica I. 28040 Madrid, Spain
}\thanks{Talk delivered on July 14th at the QCD04 meeting in Montpellier.},
Christian S. Fischer\address{IPPP, University of Durham, Durham DH1 3LE, 
U.K.}, Reinhard  Alkofer\address{Institute for Theoretical Physics, 
University of T\"ubingen, D-72076 T\"ubingen, Germany} 
} 
\vspace{0.5cm}       
\begin{document}

\begin{abstract}
We construct a model for the quark-gluon vertex of Landau gauge QCD. This
is of twofold interest: on the one hand the quark-gluon interaction
is at the heart of quark confinement, on the other hand it is a central 
element in hadron phenomenology based on QCD Greens functions.
We employ the non-Abelian one-loop diagram in perturbation theory, which 
is of order $N_c$. As a novelty we replace the
tree-level quark and gluon propagators in this diagram by their dressed
counterparts solving the Dyson-Schwinger equations.
The $N_c$-suppressed Abelian diagram is an order of magnitude
smaller in various kinematics.
We also study the effect of ghost dressing factors on the vertex obtaining 
a construction in good agreement  with recent low-momentum lattice
calculations.
\end{abstract}

% typeset front matter (including abstract)
\maketitle
\thispagestyle{empty}

%%%%%%%%%%%%%%%%%%%%%%%%%%%%%%%%%%%%%%%%%%%%%%%%%%%%%%%%%%%%%%%%%%%%%%%%%
\section{Construction of the vertex model}
%%%%%%%%%%%%%%%%%%%%%%%%%%%%%%%%%%%%%%%%%%%%%%%%%%%%%%%%%%%%%%%%%%%%%%%%%

The infrared suppression of the gluon 2-point function in QCD 
\cite{smekal,Bowman:2004jm} entails that the bare $q\bar{q}g$ vertex, 
usually employed in the rainbow truncation of the Dyson-Schwinger 
Equation (DSE), is 
insufficient to trigger dynamical chiral symmetry breaking.
Therefore we expect an infrared enhancement in the quark and gluon vertex, as 
suggested also by its Slavnov-Taylor Identity (STI). 
A model vertex with a Ball-Chiu or Curtis-Pennington structure multiplied 
by enhancing ghost factors has been successfully employed in 
\cite{fischer1} in the quark DSE. For a short summary of these results see
ref. \cite{adel}.

\begin{figure}[h]
\includegraphics[width=2in]{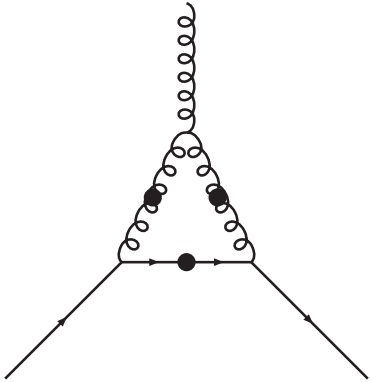}
\caption{The non-Abelian one-loop correction to the quark and gluon vertex
is the basis of our nonperturbative model. To this end we replace internal
propagators by their dressed counterparts and enhance the internal
vertices with one ghost dressing function each reflecting the STI.}
\label{NAcorrection}
\end{figure}

In this work we report a diagrammatic construction based on the one-loop 
perturbative QCD corrections to the bare vertex \cite{davydychev}. 

There are two relevant diagrams, to which we refer as Abelian and 
non-Abelian due to the vertex attached to the gluon (as is customary).
In both diagrams we substitute the quark and gluon propagators by their 
dressed counterparts solving the DSE's \cite{fischer1}. These are added to 
the bare vertex $Z_{1F}\gamma_\mu$ and the renormalization constant is 
fixed by imposing that the $\gamma_\mu$ component is unity at 
a renormalization point of $2\ GeV$. 
%This is chosen low in order to 
%compare with lattice data. A higher renormalization point shifts up the 
%$\lambda_1$ Dirac amplitude and has a small impact on others.

The non-Abelian diagram, depicted in figure \ref{NAcorrection} dominates 
over the Abelian one nominally by a factor $N_c^2$. We have checked that
1) this dominance by an order of magnitude remains after loop integration 
in the vertex for various kinematics and 2) the impact of the Abelian 
vertex on the quark DSE is smaller by the same factor by performing a 
kinematic average with the kernel of the DSE as weighting function. 
Therefore to a precision of $10\ \% 
$ or even better one can ignore the Abelian diagram.

%%%%%%%%%%%%%%%%%%%%%%%%%%%%%%%%%%%%%%%%%%%%%%%%%%%%%%%%%%%%%%%%%%%%%%%%%
\section{Numerical results}
%%%%%%%%%%%%%%%%%%%%%%%%%%%%%%%%%%%%%%%%%%%%%%%%%%%%%%%%%%%%%%%%%%%%%%%%%

The vertex is projected into the tensor basis from appendix A in  
\cite{skki}. The loop integral is calculated numerically in four 
dimensions with a standard Gauss-Legendre grid. In one computer code we 
perform the spin sums numerically, in an alternative calculation we employ 
a form code to reduce the kernel analytically to relatively complex but 
tractable scalar integrals. 

\begin{figure}[htb]
\vspace{9pt}
\epsfig{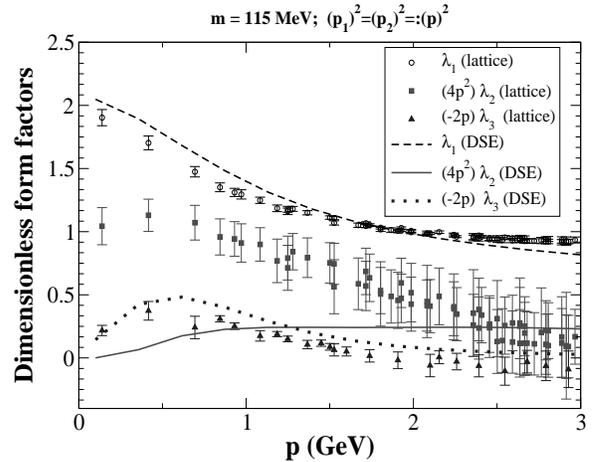}
\caption{Comparison of our results with lattice data in the particular 
kinematic section $p_1=p_2$. $m_q(2\ GeV)$ is set to $115\ MeV$ but 
similar results are obtained at the other lattice data set available at 
$60\ MeV$. }
\label{latt}
\end{figure}

If the internal $q\bar{q}g$ vertices are taken as bare $Z_{1F}\gamma_\mu$ 
then the construction, that qualitatively has the right behaviour, is not 
strong enough to reproduce lattice data and trigger chiral symmetry 
breaking in the quark DSE. This, as commented above, is expected and can 
be remedied by enhancing the internal vertices by a ghost dressing factor. 
In figure \ref{latt} we compare the resulting $q\bar{q}g$ vertex with 
lattice data at the so called ``asymmetric point'' characterized by 
$p_g=0, \ p_1=p_2$ where the gluon momentum vanishes, $p_1$ flows into 
the vertex and $p_2$ exits. This 
comparison is successful for the leading Dirac amplitude 
$\lambda_1$ and the scalar amplitude, $\lambda_3$. 
The amplitude $4 p^2 \lambda_2$ vanishes in our model at low 
momenta, whereas the lattice data (with large errors) seem to approach
a constant value implying a divergence of $\lambda_2$. 

\begin{figure}[htb]
\vspace{9pt}
\epsfig{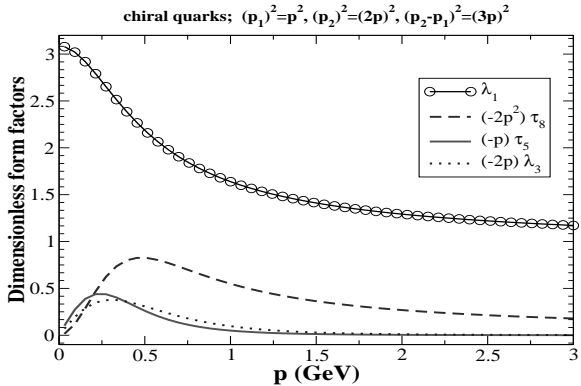}
\caption{Leading tensor structures in the kinematic section
given by $p_1:=p,\ p_2=2p,\ p_g= 3p$ in our model construction.
}
\label{asym}
\end{figure}

We now investigate a more interesting section of kinematic space, that we 
denote ``totally asymmetric'' point, characterized by the relations
$p_2=2p_1$ and $p_g= 3p_1$ between the moduli of the momenta. This point is 
interesting because the tensor basis used is  non-singular and all 
twelve different Dirac amplitudes $\lambda_{1-4}, \ \tau_{1-8}$  
contribute to the vertex. We plot the 
four leading structures in figure \ref{asym}.

The other eight Dirac amplitudes are increasingly smaller, down to two orders 
of magnitude below the plotted ones. Thus there is a rich hyrarchy of 
Dirac amplitudes that can help in model building.

%%%%%%%%%%%%%%%%%%%%%%%%%%%%%%%%%%%%%%%%%%%%%%%%%%%%%%%%%%%%%%%%%%%%%%%%%%
\section{Mass dependence and Chiral Symmetry Breaking}
%%%%%%%%%%%%%%%%%%%%%%%%%%%%%%%%%%%%%%%%%%%%%%%%%%%%%%%%%%%%%%%%%%%%%%%%%

Once we are in possession of a construction that  successfully compares
to lattice data, we employ it to perform a study of the mass dependence of
the vertex. First note the Abelian one-loop diagram contains two quark 
propagators, and is therefore suppressed as $1/M_q^2$ in the heavy quark 
limit, whereas the non-Abelian diagram will damp as $1/M_q$. Therefore 
future calculations relating observables in the charmonium and bottomonium 
systems (see \cite{roberts} for discussion) will be sensitive to what 
class of diagrams enters the vertex model.

An interesting observation is that for the range of quark masses 
considered in the lattice calculations, the $\lambda_3$ Dirac amplitude 
has a maximum.
If the current quark mass is further increased, the 
intermediate quark propagator suppresses the vertex loop. Conversely, 
approaching the chiral limit, the quark mass function takes its minimum 
value dictated by chiral symmetry breaking alone and the scalar part of 
the loop construction also has a minimum (slight corrections to this result are 
expected after self-consistently solving a vertex equation).

\begin{figure}[htb]
\vspace{9pt}
\epsfig{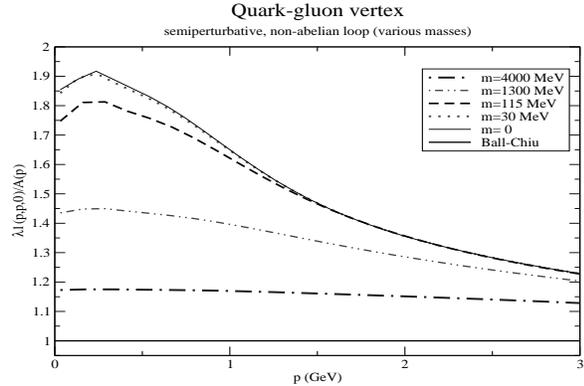}
\caption{Mass dependence of the quotient $\lambda_1/A$ that equals 1 in an
Abelian theory (as incorporated in the Ball-Chiu construction).
}
\label{massdep}
\end{figure}

As for the leading $\lambda_1$ amplitude, we plot in figure 
\ref{massdep} the quotient $\lambda_1/A$, where deviations from unity 
signal departures from the Abelian Ward-Takahashi identity.

%%%%%%%%%%%%%%%%%%%%%%%%%%%%%%%%%%%%%%%%%%%%%%%%%%%%%%%%%%%%%%%%%%%%%%%%%
\section{Outlook}
%%%%%%%%%%%%%%%%%%%%%%%%%%%%%%%%%%%%%%%%%%%%%%%%%%%%%%%%%%%%%%%%%%%%%%%%%

Results similar to ours have been independently obtained \cite{manda} in a 
different scheme. 
This exploits the STI for the three-gluon  vertex 
to model it. Since this amounts to the resummation of a totally different 
class of diagrams we would need better lattice data to distinguish both 
models. Our construction is of course valid (within approximations) for  
all possible kinematics and not just when the gluon momentum vanishes.
On the positive side, both works concur in predicting $p^2\lambda_2(p^2)$ 
to vanish as $p\to 0$. We should note that the existing lattice data, with 
large error bands, suggests instead this limit is finite, implying a 
divergence in the vertex, in the Dirac amplitude $\lambda_2$. 

When multiplied by appropriate powers of $p$ the resulting dimensionless
Dirac amplitudes of our vertex construction vanish as $p\to 0$ except 
the leading structure $\lambda_1$ that takes a finite value. 

Whether a divergence can arise as a 
consequence of the feedback of the obtained vertex model on the loop 
construction itself (implying a self-consistent solution is needed) or as 
a consequence of the backreaction on the quark SDE is a topic under 
current scrutiny.  
The construction of a quark scattering kernel based in our model 
vertex and consistent with chiral symmetry is now straight-forward.

%%%%%%%%%%%%%%%%%%%%%%%%%%%%%%%%%%%%%%%%%%%%%%%%%%%%%%%%%%%%%%%%%%%%%%%%%
\section{Questions from the audience.}
%%%%%%%%%%%%%%%%%%%%%%%%%%%%%%%%%%%%%%%%%%%%%%%%%%%%%%%%%%%%%%%%%%%%%%%%%

\begin{itemize}
\item \emph{How are Euclidean space singularities in the two-point 
functions mapped to Minkowski space after analytical 
continuation?} \\
This is a nonperturbative problem and we lack a full answer at this 
stage. One has obtained some understanding of the 
analytical structure of the two-point functions in a recent work 
\cite{maristub}. Also Hamiltonian calculations in Coulomb gauge, not in 
Euclidean space, provide clear evidence for the picture of ghost 
enhancement and gluon suppression at low momentum \cite{SwSz}. 
\item \emph{The running $\alpha_s$ you are employing seems to be somewhat 
too high in the middle-momentum range. What is the impact of this in the 
results reported?} \\
This study does not aim at precisely pinpointing 
details of the propagators and vertex functions, but to obtain insight 
into their qualitative features and structure, and identify possible 
divergences that may otherwise escape lattice calculations. In this 
respect, the impact of this few-percent deviation can be ignored.
\item \emph{Why is the dressing of the three-gluon vertex omitted?} \\
As can be observed in figure \ref{DSEvertex}, all dressing of the triple 
gluon vertex can be absorbed in the quark-gluon scattering kernel. Our 
model based on the one-loop correction to the qqg vertex can be also 
viewed as an approximation to this kernel. In this exact equation
\cite{marciano} we are neglecting completely the two last terms, 
involving the ghost-quark scattering kernel and the three-gluon-quark 
scattering kernels, as their skeleton expansion starts at two loops. 

\begin{figure}[htb]
\vspace{19pt}
\includegraphics[width=\linewidth]{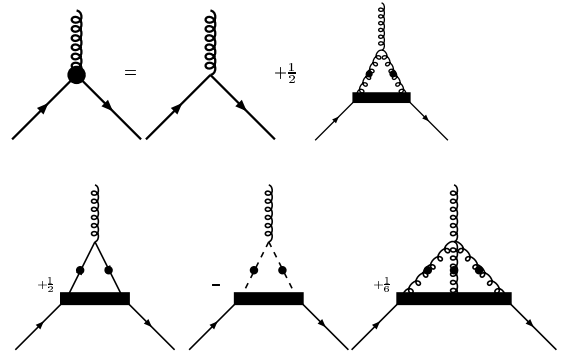}
\caption{Exact vertex DSE equation. }
\label{DSEvertex}
\end{figure}

\end{itemize}

%%%%%%%%%%%%%%%%%%%%%%%%%%%%%%%%%%%%%%%%%%%%%%%%%%%%%%%%%%%%%%%%%%%%%%%%%
\vspace{0.5cm}
\emph{We thank M. Bhagwat, C. D. Roberts and P. Tandy for valuable 
discussions. The lattice data in figure \ref{latt} are a courtesy of the 
authors of \cite{skullerud}.  F. L. E. thanks the
hospitality of the T\"ubingen Institute as well as a DAAD stipendium and 
Univ. Complutense travel grant. Partial support from grants  FPA 
2000-0956, BFM 2002-01003 (MCYT, Spain), Al 279/3-4, Fi 
970/2-1, and GRK683 (DFG, Germany).}
%%%%%%%%%%%%%%%%%%%%%%%%%%%%%%%%%%%%%%%%%%%%%%%%%%%%%%%%%%%%%%%%%%%%%%%


\begin{thebibliography}{50}
%%%%%%%%%%%%%%%%%%%%%%%%%%%%%%%%%%%%%%%%%%%%%%%%%%%%%%%%%%%%%%%%%%%%%%%

\bibitem{smekal} 
R.~Alkofer and L.~von Smekal,
%``The infrared behavior of QCD Green's functions: Confinement, dynamical
%symmetry breaking, and hadrons as relativistic bound states,''
Phys.\ Rept.\  {\bf 353} (2001) 281
[arXiv:hep-ph/0007355].
%%CITATION = HEP-PH 0007355;%%

%\cite{Bowman:2004jm}
\bibitem{Bowman:2004jm}
P.~O.~Bowman {\it et al.\/},
%``Unquenched gluon propagator in Landau gauge,''
arXiv:hep-lat/0402032.
%%CITATION = HEP-LAT 0402032;%%

\bibitem{fischer1}
C.~S.~Fischer and R.~Alkofer,
%``Non-perturbative propagators, running coupling and dynamical quark mass of
%Landau gauge QCD,''
Phys.\ Rev.\ D {\bf 67} (2003) 094020
[arXiv:hep-ph/0301094].
%%CITATION = HEP-PH 0301094;%%

\bibitem{adel}
C.~S.~Fischer, F.~Llanes-Estrada and R.~Alkofer,
arXiv:hep-ph/0407294.
%%CITATION = HEP-PH 0407294;%%

\bibitem{davydychev}
A.~I.~Davydychev, P.~Osland and L.~Saks,
%``Quark gluon vertex in arbitrary gauge and dimension,''
Phys.\ Rev.\ D {\bf 63} (2001) 014022
%%CITATION = HEP-PH 0008171;%%

\bibitem{skki}
J.~Skullerud and A.~Kizilersu,
%``Quark-gluon vertex from lattice QCD,''
JHEP {\bf 0209} (2002) 013
[arXiv:hep-ph/0205318].
%%CITATION = HEP-PH 0205318;%%

\bibitem{skullerud}
J.~I.~Skullerud, P.~O.~Bowman, A.~Kizilersu, D.~B.~Leinweber and 
A.~G.~Williams,
%``Nonperturbative structure of the quark gluon vertex,''
JHEP {\bf 0304} (2003) 047
[arXiv:hep-ph/0303176].
%%CITATION = HEP-PH 0303176;%%

\bibitem{roberts}
M.~S.~Bhagwat, A.~Holl, A.~Krassnigg, C.~D.~Roberts and P.~C.~Tandy,
%``Aspects and consequences of a dressed-quark-gluon vertex,''
arXiv:nucl-th/0403012.
%%CITATION = NUCL-TH 0403012;%%

\bibitem{manda}
M.~S.~Bhagwat and P.~C.~Tandy,
%``Quark-gluon vertex model and lattice-QCD data,''
arXiv:hep-ph/0407163.
%%CITATION = HEP-PH 0407163;%%


\bibitem{maristub}
R.~Alkofer, W.~Detmold, C.~S.~Fischer and P.~Maris,
%``Analytic properties of the Landau gauge gluon and quark propagators,''
Phys.\ Rev.\ D {\bf 70} (2004) [arXiv:hep-ph/0309077].
%%CITATION = HEP-PH 0309077;%%

\bibitem{SwSz}
A.~P.~Szczepaniak and E.~S.~Swanson,
%``Coulomb gauge QCD, confinement, and the constituent representation,''
Phys.\ Rev.\ D {\bf 65} (2002) 025012
[arXiv:hep-ph/0107078];
%%CITATION = HEP-PH 0107078;%%
%\cite{Zwanziger:2003de}
D.~Zwanziger,
%``Analytic calculation of color-Coulomb potential and color confinement,''
arXiv:hep-ph/0312254;
%%CITATION = HEP-PH 0312254;%%
%\cite{Feuchter:2004gb}
C.~Feuchter and H.~Reinhardt,
%``Quark and gluon confinement in Coulomb gauge,''
arXiv:hep-th/0402106; 
%%CITATION = HEP-TH 0402106;%%
\emph{id.} work in preparation.

\bibitem{marciano}
W.~J.~Marciano and H.~Pagels,
%``Quantum Chromodynamics: A Review,''
Phys.\ Rept.\  {\bf 36} (1978) 137;
%%CITATION = PRPLC,36,137;%%
E.~Eichten and F.~Feinberg,
%``Dynamical Symmetry Breaking Of Nonabelian Gauge Symmetries,''
Phys.\ Rev.\ D {\bf 10} (1974) 3254.
%%CITATION = PHRVA,D10,3254;%%


\end{thebibliography}
\end{document}